\documentclass[prc,aps,floats,twocolumn,twoside,preprintnumbers,
superscriptaddress,floatfix,nofootinbib,showpacs]{revtex4}
\usepackage{graphicx,epsfig,pstricks,rotating}
\usepackage{bm}

\renewcommand{\deg}{^{\circ}}

\newcommand{\pipsq}[1]{\left[ #1 \right]} 

\newcommand{\eqnb}{\begin{equation}}
\newcommand{\eqne}{\end{equation}}

\begin{document}


\let\I\i
\def\i{\mathrm{i}}
\def\e{\mathrm{e}}
\def\d{\mathrm{d}}
\def\half{{\textstyle{1\over2}}}
\def\thalf{{\textstyle{3\over2}}}
\def\h{{\scriptscriptstyle{1\over2}}}
\def\th{{\scriptscriptstyle{3\over2}}}
\def\fh{{\scriptscriptstyle{5\over2}}}
\def\vec#1{\mbox{\boldmath$#1$}}
\def\svec#1{\mbox{{\scriptsize \boldmath$#1$}}}
\def\oN{\overline{N}}
\def\ttimes{{\scriptstyle \times}}
\def\bm#1{{\pmb{\mbox{${#1}$}}}}

\def\CG#1#2#3#4#5#6{C^{#5#6}_{#1#2#3#4}}
\def\threej#1#2#3#4#5#6{\left(\begin{array}{ccc}
    #1&#2&#3\\#4&#5&#6\end{array}\right)}

\title{$\eta$ and $\eta'$ mesons in the Dyson-Schwinger approach using a generalization of the Witten-Veneziano relation}
\author{D. Horvati\' c}
\email{davorh@phy.hr}
\affiliation{Physics Department, Faculty of Science,
University of Zagreb,
Bijeni\v{c}ka c. 32, 10000 Zagreb, Croatia}
\author{D. Blaschke}
\email{blaschke@ift.uni.wroc.pl}
\affiliation{Institute for Theoretical Physics, University of Wroclaw, 
Max Born pl. 9, 50-204 Wroclaw, Poland }
\affiliation{ Bogoliubov  Laboratory of Theoretical Physics,
Joint Institute for Nuclear Research, 141980 Dubna, Russia}
\affiliation{
Institute of Physics, University of Rostock,
D-18051 Rostock, Germany}
\author{Yu. Kalinovsky}
\email{kalinov@jinr.ru}
\affiliation{Laboratory for Information Technologies,
Joint Institute for Nuclear Research, 141980 Dubna, Russia}
\author{D. Kekez}
\email{kekez@lei.irb.hr}
\affiliation{Rudjer Bo\v{s}kovi\'{c} Institute, P.O.B. 180,
10002 Zagreb, Croatia}
\author{D. Klabu\v{c}ar}
\email{klabucar@phy.hr}
\thanks{Corresponding author}
\thanks{Senior Associate of Abdus Salam ICTP, Trieste, Italy}
\affiliation{Physics Department, Faculty of Science,
University of Zagreb, Bijeni\v{c}ka c. 32, 10000 Zagreb, Croatia}

\begin{abstract}
The description of the $\eta$ and $\eta^\prime$ mesons in the 
Dyson-Schwinger approach has relied on the Witten-Veneziano 
relation. The present paper explores the consequences of using instead 
its generalization recently proposed by Shore. 
On the examples of three different model interactions,
we find that irrespective of the concrete model dynamics, our 
Dyson-Schwinger approach is phenomenologically more successful in
conjunction with the standard Witten-Veneziano relation than with 
the proposed generalization valid in all orders in the $1/N_c$ expansion.
\end{abstract}
\pacs{11.10.St, 12.38.-t, 14.40.Aq, 12.38.Lg}

\maketitle

\section{Introduction}
\label{INTRO}

\noindent
The Dyson-Schwinger (DS) approach 
\cite{ForExample,Alkofer:2000wg,Roberts:2000hi,Roberts:2000aa,Holl:2006ni,Fischer:2006ub} 
to QCD and its modeling is the chirally well-behaved bound-state approach.  Thus, it is 
the most suitable bound-state approach to treat the light pseudoscalar mesons 
(those composed of the $u, d$ and $s$ quarks), for which 
dynamical chiral symmetry breaking (DChSB) is essential.
One solves the (``gap") DS equations (DSEs) 
${S}_q^{-1} = (S^{\scriptstyle {\rm free}}_{q})^{-1} - \Sigma_q$ 
for the dynamically dressed quark propagators 
${S}_q$, where $q$ is the quark flavor ($q=u,d,s, ...$),
$S^{\scriptstyle {\rm free}}_q$ is the {free} quark propagator, and 
$\Sigma_q$ is the quark self-energy.
These dressed quark propagator solutions are then employed in 
Bethe-Salpeter equations (BSEs)
for the bound-state vertex $\Gamma_{q{\bar q}'}$ of
the meson composed of the quark of the flavor $q$ and antiquark
of the flavor $q'$:
\begin{equation}
[\Gamma_{q{\bar q}'}]_{ef} = \int [S_q \Gamma_{q{\bar q}'} S_{q'} ]_{gh} 
[K]_{ef}^{hg} \, \, ,
\label{BSE}
\end{equation}
where $e,f,g,h$ {\it schematically} represent spinor, color
and flavor indices, integration is meant over loop momenta,
and $K$ is the interaction kernel. Solving Eq. (\ref{BSE}) 
for $\Gamma_{q{\bar q}'}$ also yields $M_{q\bar q'}$, 
the mass eigenvalue of the $q{\bar q}'$ meson.

To obtain the chiral behavior as in QCD, DS and BS equations
must be solved in a consistent approximation. The rainbow-ladder 
approximation (RLA), where DChSB is well-understood, is still 
the most usual approximation in phenomenological applications.
This also entails that in both DSE and BSE (\ref{BSE}) we employ 
the same effective interaction kernel, 
\begin{equation}
[K(k)]_{ef}^{hg} = {\rm i} \,
      g^2 D_{\mu\nu}^{ab}(k)_{\mbox{\rm\scriptsize eff}} \,
[\frac{\lambda^a}{2}\,\gamma^{\mu}]_{eg} \,
[\frac{\lambda^a}{2}\,\gamma^{\nu}]_{hf} \, ,
\label{RLAkernel}
\end{equation}
so that the quark self energy in the gap DSE is
\begin{eqnarray}
 \label{DS-equation}
\Sigma_q(p) =  
- \int \!\!\frac{d^4\ell}{(2\pi)^4} \,
  g^2 D_{\mu\nu}^{ab}(k)_{\mbox{\rm\scriptsize eff}} \, 
\frac{\lambda^a}{2}\,\gamma^{\mu}
S_q(\ell) \frac{\lambda^b}{2}\,\gamma^{\nu}~. 
\end{eqnarray}
In Eqs. (\ref{RLAkernel}) and (\ref{DS-equation}), 
$D_{\mu\nu}^{ab}(k)_{\mbox{\rm\scriptsize eff}}$ is an effective gluon 
propagator. For example, for renormalization-group improved (RGI) 
interactions (e.g., in Refs. 
\cite{jain93b,Klabucar:1997zi,Kekez:2000aw,Kekez:2003ri,Kekez:2005ie}), 
it has the form 
\begin{equation}
g^2 D_{\mu\nu}^{ab}(k)_{\mbox{\rm\scriptsize eff}}
= 
4\pi\alpha_{\mbox{\rm\scriptsize eff}}(k^2)
D_{\mu\nu}^{ab}(k)_{\scriptstyle {\rm free}}
\end{equation}
where $D_{\mu\nu}^{ab}(p)_{\scriptstyle {\rm free}}$ is the free gluon propagator,
and $\alpha_{\mbox{\rm\scriptsize eff}}(k^2)$ is an {\it effective} running 
coupling. For large spacelike momenta ($k^2 \gg 1$ GeV$^2$), 
$\alpha_{\mbox{\rm\scriptsize eff}}(k^2)$ approaches the perturbative 
QCD running coupling $\alpha_{\mbox{\rm\scriptsize s}}(k^2)$ known from the
QCD renormalization group analysis, although it must be modeled at
low momenta.

Concretely, in the present paper we recall and utilize the results obtained 
{\it i)} in Refs. \cite{Klabucar:1997zi,Kekez:2000aw} by using the  
RGI of Jain and Munczek \cite{jain93b}, {\it ii)} in Ref. \cite{Kekez:2005ie} 
by using the RGI gluon condensate-induced interaction \cite{Kekez:2003ri}, 
and {\it iii)} in Refs. \cite{Horvatic:2007wu,Horvatic:2007qs} by using 
the separable interaction \cite{Blaschke:2000gd}. In any case,
such effective interactions must be modeled at least in the low-energy, 
nonperturbative regime in order to be phenomenologically successful --
which above all means to be sufficiently strong in the low-momentum 
domain to yield DChSB. 
In the chiral limit (and {\it close} to it), light pseudoscalar ($P$) 
meson $q\bar q$ bound states ($P=\pi^{0,\pm}, K^{0,\pm}, \eta$) 
then {simultaneously} manifest themselves also as 
\mbox{({\it quasi-})}Goldstone bosons of DChSB.
This enables one to work with the mesons as 
explicit $q\bar q$ bound states, 
while reproducing 
the results of the Abelian axial anomaly for the light pseudoscalars, 
i.e., the amplitudes for $P\rightarrow\gamma\gamma$ and 
$\gamma^\star \rightarrow P^0 P^+ P^-$.
This is unique among the bound state approaches -- e.g., see
Refs. \cite{Roberts:2000hi,Kekez:1998xr,Alkofer:1995jx,Bistrovic:1999dy} 
and references therein. Nevertheless, one keeps the advantage of bound-state 
approaches that from the $q\bar q$ substructure one can calculate many 
important quantities (such as the pion, kaon and $s\bar s$ pseudoscalar 
decay constants: $f_\pi$, $f_K$ and $f_{s\bar s}$) which are just parameters 
in most of other chiral approaches to the light-quark sector. The treatment 
\cite{Klabucar:1997zi,Kekez:2000aw,Kekez:2001ph,Kekez:2005ie}
of the $\eta$-$\eta'$ complex is remarkable in that it is 
very successful in spite of the limitations of RLA.
(Very recently, during the work on the present paper,
the first and still simplified DS treatments of $\eta$ and 
$\eta'$ beyond RLA appeared \cite{Lakhina:2007vm,Bhagwat:2007ha}. 
However, RLA treatments 
will probably long retain their usefulness in applications
where simple modeling is desirable, as in the calculationally
demanding finite-temperature calculations \cite{Horvatic:2007qs}.)
The RLA treatments of the $\eta$-$\eta'$ complex at first determined 
\cite{Klabucar:1997zi,Kekez:2000aw,Kekez:2001ph} the anomalous $\eta_0$ mass 
parameter by fitting the empirical $\eta$ and $\eta'$ masses. More recently,
the treatment was improved by avoiding this fitting while retaining the 
phenomenologically successful description \cite{Kekez:2005ie,Horvatic:2007qs}. 
Namely, the anomalous $\eta_0$ mass was no longer a free parameter 
but determined from the lattice results (on QCD topological susceptibility) 
through the Witten-Veneziano (WV) relation
\cite{Witten:1979vv,Veneziano:1979ec}. However, Shore achieved
\cite{Shore:2006mm,Shore:2007yn} what can be considered as a
generalization of the WV relation, and the purpose of the
present paper is exploring the usage of
this generalization in the DS context.

The paper is organized as follows: in the next section, we recapitulate 
the procedures and results of our previous treatments 
\cite{Kekez:2000aw,Kekez:2005ie,Horvatic:2007qs} relying on 
the WV relation (\ref{WittenVenez}), and present in Table I also 
their extension to the scheme of the four decay constants (and two 
mixing angles) of $\eta$ and $\eta'$. 
In Section \ref{ShoresRelations}, we expose the usage of the
pertinent Shore's equations \cite{Shore:2006mm,Shore:2007yn}
in the context of DS approach.  The last section concludes 
after giving the results of solving the pertinent equations.

\section{$\eta$-$\eta'$ mass matrix from Witten-Veneziano relation}
\label{massMatrixAndWVrelation}

All $q\bar q'$ model masses $M_{q\bar q'}$ ($q, q' = u,d,s$)
used in the present paper, and corresponding $q\bar q'$ bound-state 
amplitudes, were obtained in Refs. 
\cite{Klabucar:1997zi,Kekez:2000aw,Kekez:2005ie,Blaschke:2007ce,Horvatic:2007wu,Horvatic:2007qs}
in RLA, i.e., 
with an interaction kernel which 
(irrespective of how one models the dynamics) cannot possibly 
capture the effects of the non-Abelian, gluon axial anomaly. 
Thus, when we form the $\eta$-$\eta'$ mass matrix 
\begin{equation}
{\hat M}^2_\mathrm{NA} =
\left[ \begin{array}{cl}  M_{88}^2 & M_{80}^2\\
                          M_{08}^2 & M_{00}^2
        \end{array} \right]~,
\label{M2NA}
\end{equation}
in this case in the octet-singlet basis $\eta_8$-$\eta_0$
of the (broken) flavor-SU(3) states of isospin zero, 
\begin{equation}
  \eta_8
        =
  \frac{1}{\sqrt{6}}(u\bar{u} + d\bar{d} -2 s\bar{s}),
\quad         
  \eta_0
        =
  \frac{1}{\sqrt{3}}(u\bar{u} + d\bar{d} + s\bar{s}),
\label{etasdef}
        \end{equation}
this matrix (\ref{M2NA}), consisting of our calculated $q\bar q$ masses,
\begin{equation}
M_{88}^2 \equiv \langle \eta_8 | {\hat M}^2_\mathrm{NA} |\eta_8 \rangle
 = \frac{2}{3}\, (M_{s\bar{s}}^2 + \frac{1}{2}M_{u\bar{u}}^2) \, ,
\end{equation}
\begin{equation}
M_{80}^2 \equiv \langle \eta_8 | {\hat M}^2_\mathrm{NA} |\eta_0 \rangle
= M_{08}^2 = \frac{\sqrt{2}}{3} ( M_{u\bar{u}}^2 - M_{s\bar{s}}^2 )
< 0,
\end{equation}
\begin{equation}
M_{00}^2 \equiv \langle \eta_0| {\hat M}^2_\mathrm{NA} |\eta_0 \rangle
= \frac{2}{3}\, (\frac{1}{2}M_{s\bar{s}}^2 + M_{u\bar{u}}^2) \, ,
\end{equation}
is purely non-anomalous ($\mathrm{NA}$), vanishing in the chiral limit. In 
the isospin limit, to which we adhere throughout, the pion is strictly 
decoupled from the gluon anomaly and $M_{u\bar{u}} = M_{d\bar{d}}$ is 
exactly our model pion mass $M_{\pi}$.
Also the unphysical $s\bar s$ quasi-Goldstone's mass $M_{s\bar s}$ 
results from RLA BSE and does not include the contribution from the 
gluon anomaly. This is consistent with the fact that due to the 
Dashen-Gell-Mann-Oakes-Renner (DGMOR) relation, it is in a good 
approximation 
\cite{Klabucar:1997zi,Kekez:2000aw,Kekez:2005ie,Horvatic:2007qs} 
given by
\begin{equation}
M_{s\bar s}^2 = 2 M_K^2 - M_{\pi}^2 \, ,
\label{MssbarMKMpi}
\end{equation}
i.e., by the kaon and pion masses protected from the anomaly
by strangeness and/or isospin. 

In our previous DS studies 
\cite{Klabucar:1997zi,Kekez:2000aw,Kekez:2005ie,Blaschke:2007ce,Horvatic:2007wu,Horvatic:2007qs},
to which we refer for all model details, the phenomenology of the
non-anomalous sector was successfully reproduced, e.g., $f_\pi$, 
$f_K$, as well as the empirical masses $M_{\pi}$ and 
$M_K$ (see the upper part of Table \ref{3WV+Shore}), 
yielding a strongly non-diagonal ${\hat M}^2_\mathrm{NA}$ (\ref{M2NA}). 
Its diagonalization leads to the eigenstates known as the 
nonstrange-strange ($\mathrm{NS}$-$\mathrm{S}$) basis,
\begin{equation}
\eta_\mathrm{NS} = \frac{1}{\sqrt{2}} (u\bar{u} + d\bar{d})~,
\qquad
\eta_\mathrm{S} = s\bar{s} \, ,
\label{NS-Sbasis}
\end{equation}
and to 
${\hat M}^2_\mathrm{NA}={\rm diag}[M_{\pi}^2,M_{s\bar s}^2]$. 
In contrast to these mass-squared eigenvalues, the experimental 
masses are such that $(M_{\pi}^2)_{exp} \ll (M_{\eta}^2)_{exp}$, 
and $\eta'$ is too heavy, $(M_{\eta'})_{exp} = 958$ MeV,
to be considered even as the $s\bar s$ quasi-Goldstone boson.
This is the well-known $U_A(1)$ problem, resolved by the
fact that the {\it complete} $\eta$-$\eta'$ mass matrix 
${\hat M}^2$ must contain the anomalous ($A$) part ${\hat M}^2_\mathrm{A}$.
That is, ${\hat M}^2 = {\hat M}^2_\mathrm{NA} +  {\hat M}^2_\mathrm{A}$. 

However, ${\hat M}^2_\mathrm{A}$ is inaccessible to RLA
which yields our Goldstone pseudoscalars. In Refs. 
\cite{Klabucar:1997zi,Kekez:2000aw,Kekez:2005ie,Horvatic:2007wu,Horvatic:2007qs},
${\hat M}^2_\mathrm{A}$ was extracted from lattice data through the WV relation
[the second equality in Eq. (\ref{WittenVenez})]. 
The purpose of the present paper, instead, is to approach 
$\eta$ and $\eta'$ through Shore's \cite{Shore:2006mm,Shore:2007yn}
recent generalization of that relation. 

Before that, however, we review the usage of the WV relation in Refs.
\cite{Klabucar:1997zi,Kekez:2000aw,Kekez:2005ie,Horvatic:2007wu,Horvatic:2007qs}.
The expansion in the large number of colors, $N_c$, indicates that 
the leading approximation in that expansion describes the bulk
of main features of QCD.
The gluon anomaly is suppressed as $1/N_c$ and can be viewed as a 
perturbation in the large $N_c$ expansion. In the SU(3) limit
[compare Eqs. (\ref{M2A}) and (\ref{M2AqqX})], it is coupled {\it only} 
to the singlet combination $\eta_0$ (\ref{etasdef}); only the $\eta_0$ 
mass receives, from the gluon anomaly, a contribution which, unlike 
quasi-Goldstone masses $M_{q\bar q'}$'s comprising ${\hat M}^2_\mathrm{NA}$, 
does {\it not} vanish in the chiral limit.
As discussed in Refs. \cite{Klabucar:1997zi,Kekez:2005ie}, 
in the present bound-state context it is thus meaningful to 
include the effect of the gluon anomaly just on the level of
a mass shift for the $\eta_0$ as the lowest-order effect, and retain
the $q\bar q$ bound-state amplitudes and the corresponding 
mass eigenvalues $M_{q\bar q}$ as calculated by solving DSEs and
BSEs with kernels in RLA. 

References 
\cite{Klabucar:1997zi,Kekez:2000aw,Kekez:2005ie,Horvatic:2007wu,Horvatic:2007qs}
thus break the $U_A(1)$
symmetry, and avoid the $U_A(1)$ problem, by shifting the $\eta_0$ 
(squared) mass by an amount denoted by $3\beta$ (in the notation of 
Refs. \cite{Kekez:2000aw,Kekez:2005ie}). The complete mass matrix 
${\hat M}^2 = {\hat M}^2_\mathrm{NA} +  {\hat M}^2_\mathrm{A}$ then contains
the anomalous part
\begin{equation}
{\hat M}^2_\mathrm{A} = \mbox{\rm diag}[0, 3\beta]~,
\label{M2A}
\end{equation}
where the anomalous $\eta_0$ mass shift $3\beta$ 
is related to the topological susceptibility of the vacuum,
but in the present approach must be treated as a parameter to 
be determined outside of our RLA model, i.e., fixed by
phenomenology or taken from the lattice calculations \cite{Alles:1996nm}.
(The possibility of employing an additional microscopic model for the 
gluon anomaly contribution, such as the one of Ref. \cite{Blaschke:1996dp},
is presently not considered.)

The SU(3) flavor symmetry breaking and its interplay with the
gluon anomaly modifies \cite{Kekez:2005ie} 
${\hat M}^2_\mathrm{A}$ (\ref{M2A}) to 
\begin{equation}
{\hat M}^2_\mathrm{A} = 
\beta \left[ \begin{array}{cl}
      \,\,  \frac{2}{3}(1-X)^2 & \frac{\sqrt{2}}{3}(1-X)(2+X) \\
  \frac{\sqrt{2}}{3}(1-X)(2+X) & \,\,\,\,\, \quad \frac{1}{3}(2+X)^2
        \end{array} \right]
 \, ~,
\label{M2AqqX}
\end{equation}
where $X$ is the flavor symmetry breaking parameter. It is most often
estimated as $X = f_\pi/f_{s\bar s} \sim 0.7-0.8$ (see, e.g., Refs. 
\cite{FeldmannKrollStech98PRD,Feldmann99IJMPA,Kekez:2000aw,Kekez:2005ie},
although there are some other \cite{Kekez:2000aw}, of course related, 
estimates of $X$). Presently we also adopt $X = f_\pi/f_{s\bar s}$, 
which means that $X$ is a calculated quantity in our approach.
The employed models achieved good agreement with phenomenology 
\cite{Klabucar:1997zi,Kekez:2000aw,Kekez:2005ie,Horvatic:2007qs}, 
e.g., fitted the experimental 
value of $M_\eta^2 + M_{\eta'}^2$ for $\beta$ around 0.26 -- 0.28 GeV$^2$.
The anomaly contribution ${\hat M}^2_\mathrm{A}$ then brings the complete $M^2$ 
rather close to a diagonal form for all considered models
\cite{Klabucar:1997zi,Kekez:2000aw,Kekez:2005ie,Horvatic:2007qs}; 
that is, to diagonalize $M^2$, only a relatively small rotation 
($|\theta|\sim 13^\circ \pm 2^\circ$) 
of the $\eta_{8}$-$\eta_{0}$ basis states,
\begin{equation}
\eta = \cos\theta \, \eta_{8}
             - \sin\theta \, \eta_0~,
\,\,\,\,\,\,\,
\eta^\prime = \sin\theta \, \eta_8
             + \cos\theta \, \eta_0~,
\label{MIXtheta}
\end{equation}
is needed to align them with the mass eigenstates, i.e.,
with the physical $\eta$ and $\eta^\prime$.
In contrast to this,
the $\eta$-$\eta^\prime$ mass matrix in the $\mathrm{NS}$-$\mathrm{S}$
basis (\ref{NS-Sbasis}),
\begin{eqnarray}
\label{M2_NS-S}
{\hat M}^2 &=&
             \pipsq{
                \begin{array}{ll}
         M_{\eta_{\mathrm{NS}}}^2    &  M_{\eta_{\mathrm{S}}\eta_{\mathrm{NS}}}^2 \\
            M_{\eta_{\mathrm{NS}}\eta_{\mathrm{S}}}^2 &    M_{\eta_{\mathrm{S}}}^2
                \end{array}
        }  \\
     &=&
       \pipsq{
                \begin{array}{ll}
      M_{\pi}^2 + 2 \beta  & \quad \sqrt{2} \beta X \\
        \,  \sqrt{2} \beta X    & M_{s\bar{s}}^2 + \beta X^2
                \end{array}
        }
\begin{array}{c} \vspace{-2mm} \longrightarrow \\ \phi \end{array}
        \pipsq{
                \begin{array}{ll}
                        M_\eta^2        & \,\,  0 \\
                     \,   0               & M_{\eta'}^2
                \end{array}
        },
\nonumber
\end{eqnarray}
is then strongly off-diagonal.
The indicated diagonalization, given by
\begin{equation}
\eta = \cos\phi \, \eta_{\mathrm{NS}}
             - \sin\phi \, \eta_\mathrm{S}~,
\,\,\,\,\,\,\,
\eta^\prime = \sin\phi \, \eta_\mathrm{NS}
             + \cos\phi \, \eta_\mathrm{S}~,
\label{MIXphi}
\end{equation}
is thus achieved for a large $\mathrm{NS}$-$\mathrm{S}$ state-mixing angle
$\phi \sim 42^\circ \pm 2^\circ$. Of course, this is again in agreement 
with phenomenological requirements \cite{Kekez:2000aw,Kekez:2005ie}, 
since $\phi$ is fixed to the $\eta_8$-$\eta_0$ state-mixing angle 
$\theta$ by the relation 
$\phi = \theta + \arctan\sqrt{2}  = \theta + 54.74\deg$.
The masses are
\begin{eqnarray}
\label{Meta}
\nonumber
M_{\eta}^2 &=& \frac{1}{2} \left[ M_{\eta_{\mathrm{NS}}}^2 + M_{\eta_{\mathrm{S}}}^2
  - \sqrt{(M_{\eta_{\mathrm{NS}}}^2  - M_{\eta_{\mathrm{S}}}^2)^2 + 8 \beta^2 X^2} \right],
\\
M_{\eta'}^2 &=& \frac{1}{2} \left[ M_{\eta_{\mathrm{NS}}}^2 + M_{\eta_{\mathrm{S}}}^2
  + \sqrt{(M_{\eta_{\mathrm{NS}}}^2  - M_{\eta_{\mathrm{S}}}^2)^2 + 8 \beta^2 X^2} \right].
\nonumber
\label{MetaPrime}
\end{eqnarray}

The invariant trace of the mass matrix (\ref{M2_NS-S}), together
with Eq. (\ref{MssbarMKMpi}), 
gives the first equality in 
\begin{equation}
   \beta \, (2 + X^2) = M_\eta^2 + M_{\eta'}^2 - 2 M_K^2 =
\frac{6}{f_\pi^2} \, \chi_{\rm YM}
 \, .
\label{WittenVenez}
\end{equation}
The second equality is the Witten-Veneziano (WV) relation 
\cite{Witten:1979vv,Veneziano:1979ec} between the $\eta$, 
$\eta'$ and kaon masses and $\chi_{\rm YM}$, the topological 
susceptibility of the pure gauge, Yang-Mills theory.
Thus, $\beta$ does not need to be a free parameter, but can be
determined from lattice results on $\chi_{\rm YM}$, so that 
no fitting parameters are introduced. For the three models
\cite{jain93b,Kekez:2003ri,Blaschke:2000gd} utilized in our treatments  
\cite{Klabucar:1997zi,Kekez:2000aw,Kekez:2005ie,Horvatic:2007qs} of
$\eta$ and $\eta'$, the bare quark mass parameters and the 
interaction parameters were fixed already in the non-anomalous sector, 
by requiring the good pion and kaon phenomenology.
(See the $\pi$ and $K$ masses and decay constants in
the uppermost part of Table \ref{3WV+Shore}.)
Then, following Refs. \cite{Kekez:2005ie,Horvatic:2007qs} in
adopting the central value of the weighted average of the recent 
lattice results on Yang-Mills topological susceptibility 
\cite{Lucini:2004yh,DelDebbio:2004ns,Alles:2004vi}, 
\begin{equation}
\chi_{\rm YM} = (175.7 \pm 1.5 \, \rm MeV)^4 \, ,
\label{chiYMaverage}
\end{equation}
we have obtained 
the good descriptions of the $\eta$-$\eta'$ phenomenology
\cite{Klabucar:1997zi,Kekez:2000aw,Kekez:2005ie,Horvatic:2007qs}, 
exemplified by the first three columns (one for each DS models used) 
of the middle part of Table \ref{3WV+Shore}, giving the predictions for 
the $\eta$ and $\eta'$ masses and for the $\mathrm{NS}$-$\mathrm{S}$ mixing angle $\phi$.
 
The lowest part of the table, below the second horizontal dividing line, 
contains the results on the quantities ($\theta_0$, $\theta_8$, etc.) 
defined in the scheme with four $\eta$ and $\eta^\prime$ decay constants 
and two mixing angles, introduced and explained in the following 
Section~\ref{ShoresRelations}.
Table \ref{3WV+Shore} also compares
these results of ours (in the first three columns) with the corresponding 
results of Shore's approach~\cite{Shore:2006mm,Shore:2007yn}, in which 
the {\it experimental} values of the meson masses $M_\pi$, $M_K$, $M_\eta$, 
and $M_{\eta^\prime}$, as well as the decay constants $f_\pi$ and $f_K$ 
(in contrast to our $q\bar q$ bound-state model predictions for these 
quantities) are used as inputs enabling the calculation of various decay 
constants in the $\eta$-$\eta'$ complex and the two mixing angles $\theta_0$ 
and $\theta_8$ (corresponding to $\phi = 38.24^\circ$ in our approach).

\begin{table}[b]
\begin{center}
\begin{tabular}{|c|c|c|c|c|c|}
\hline
from & \cite{Kekez:2000aw} & \cite{Kekez:2005ie}     & \cite{Horvatic:2007qs}     & Shore                            &           \\
Ref. &              \& WV &                 \& WV &                    \& WV & \cite{Shore:2006mm,Shore:2007yn} & Experiment \\
\hline
$M_\pi$            & 137.3   & 135.0      & 140.0   &    & $(138.0)^{\mathrm{isospin}}_{\mathrm{average}}$ \\
$M_K$              & 495.7   & 494.9      & 495.0   &    & $(495.7)^{\mathrm{isospin}}_{\mathrm{average}}$\\
$M_{s\bar s}$      & 700.7   & 722.1      & 684.8   &           &                  \\
$f_\pi$            & 93.1    & 92.9       & 92.0    &           & $92.4 \pm 0.3 $ \\
$f_K$              & 113.4   & 111.5      & 110.1   &           & $113.0 \pm 1.0$  \\
$f_{s\bar s}$      & 135.0   & 132.9      & 119.1   &           &                   \\
\hline
$M_\eta$           & 568.2      & 577.1      & 542.3      &           & $547.75\pm 0.12$ \\
$M_{\eta^\prime}$  & 920.4      & 932.0      & 932.6      &           & $957.78\pm 0.14$ \\
$\phi$           & $41.42^\circ$  & $39.56^\circ$  & $40.75^\circ$  & $(38.24^\circ)$ &                  \\
\hline
$\theta_0$         & $-2.86^\circ$     & $-5.12^\circ$     & $-6.80^\circ$     & $-12.3^\circ$     & \\
$\theta_8$         & $-22.59^\circ$    & $-24.14^\circ$    & $-20.58^\circ$    & $-20.1^\circ$     & \\
$f_0$              & 108.8         & 107.9         & 101.8         & 106.6         & \\
$f_8$              & 122.6         & 121.1         & 110.7         & 104.8         & \\
$f_\eta^0$         & 5.4           & 9.6           & 12.1          & 22.8          & \\
$f_{\eta^\prime}^0$& 108.7         & 107.5         & 101.1         & 104.2         & \\
$f_\eta^8$         & 113.2         & 110.5         & 103.7         & 98.4          & \\
$f_{\eta^\prime}^8$& -47.1         & -49.5         & -38.9         & -37.6         & \\
\hline
\end{tabular}
\end{center}
\caption{The results of employing the WV relation (\ref{WittenVenez}) 
in our DS approach for the three dynamical models used in Refs. 
\cite{Kekez:2000aw,Kekez:2005ie,Horvatic:2007qs},
compared with the results of Shore's analysis \cite{Shore:2006mm,Shore:2007yn}
and with the experimental results.
The first column was obtained by the WV-recalculation of the results of 
Ref. \cite{Kekez:2000aw}, which in turn used the Jain-Munczek 
{\em Ansatz} for the gluon propagator \cite{jain93b}.
Column 2: the results based on Ref. \cite{Kekez:2005ie}, which used 
the OPE-inspired, gluon-condensate-enhanced gluon propagator \cite{Kekez:2003ri}.
Column  3: the results based on Ref. \cite{Horvatic:2007qs}, which utilized the 
separable {\em Ansatz} for the dressed gluon propagator \cite{Blaschke:2000gd}.
Column 4: The results of Shore \cite{Shore:2006mm,Shore:2007yn}, who used the 
lattice result $\chi_{\mbox{\rm\scriptsize YM}} = (191\, \rm MeV)^4$ of 
Ref.~\cite{DelDebbio:2004ns}, and not the weighted average (\ref{chiYMaverage}),
in contrast to us.
Column 5: the experimental values. All masses and decay constants are in 
MeV, and angles are in degrees. For more details, see text.
}
\label{3WV+Shore}
\end{table}

\section{Usage of Shore's equations in DS approach}
\label{ShoresRelations}

The WV relation was derived in the lowest-order approximation
in the large $N_c$ expansion. 
However, considerations by Shore \cite{Shore:2006mm,Shore:2007yn}
contain what amounts to the generalization of the WV relation, 
which is valid to all orders in $1/N_c$. Among the relations he
derived through the inclusion of the gluon anomaly in DGMOR relations, 
the following are pertinent for the present paper:
\begin{eqnarray}
\!\! (f^0_{\eta'})^2 M_{\eta'}^2 + (f^0_{\eta})^2 M_\eta^2 =
{1\over3} \bigl(f_\pi^2 M_\pi^2 + 2 f_K^2 M_K^2\bigr) + 6 A~, 
\quad
\label{eq:bn}\\
\nonumber\\
\!\! f^0_{\eta'} f^8_{\eta'} M_{\eta'}^2 + f^0_{\eta} f^8_{\eta} M_{\eta}^2 =
{2\sqrt2\over3}\bigl(f_\pi^2 M_\pi^2 - f_K^2 M_K^2\bigr)~,  \qquad
\label{eq:bo}\\
\nonumber\\
\!\! (f^8_{\eta'})^2 M_{\eta'}^2 + (f^8_{\eta})^2 M_{\eta}^2 =
-{1\over3}\bigl(f_\pi^2 M_\pi^2 - 4 f_K^2 M_K^2\bigr)~,    \qquad
\label{eq:bp}
\end{eqnarray}
where $A$ is the full QCD topological charge parameter, and
$f^0_{\eta'}, f^0_{\eta}, f^8_{\eta'}, f^8_{\eta}$ are the {\it four}
decay constants \cite{Gasser:1984gg,Leutwyler98,KaiserLeutwyler98}
associated with the two isoscalar pseudoscalars $\eta$ and $\eta'$.

The nonperturbative parameter $A$ is related to the
QCD topological susceptibility, quark condensates and
quark masses \cite{Shore:2006mm,Shore:2007yn}.
At large $N_c$, it should be  well-approximated by the topological susceptibility,
$A \approx \chi$. More precisely, it reduces to the YM topological susceptibility
in the large $N_c$ limit: $A = \chi_{\rm YM} + {\cal O}({1}/{N_c})$,
but at present it is not known better than that, as there are still
no lattice data on this nonperturbative QCD parameter.
Therefore, in his own phenomenological analysis,
Shore himself had to approximate $A$ by a value of
$\chi_{\rm YM}$ \cite{Shore:2006mm,Shore:2007yn}. In that sense,
because of this crucial assumption based on the lowest-order ${1}/{N_c}$
approximation, even his analysis was not (and, because of the lack of the
corresponding lattice data, could not be) carried out {\it numerically}
consistently in the orders of ${N_c}$, even though
his {\it formulas} are valid in all orders in the ${1}/{N_c}$ expansion.

While the present bound-state DS approach clearly cannot improve on the
consistency aspect, it offers the possibility of a phenomenological analysis
entirely different from Shore's.
Namely, in addition to $A \approx \chi_{\rm YM}$, Shore used
the experimentally known quantities (pion, kaon, $\eta$ and $\eta'$ masses,
as well as the pion and kaon decay constants) as inputs in
Eqs. (\ref{eq:bn})-(\ref{eq:bp}) to obtain the $\eta$ and $\eta'$ decay
constants $f^0_{\eta'}, f^0_{\eta}, f^8_{\eta'}, f^8_{\eta}$.
On the other hand, the predicting power of our bound-state DS approach is
much larger: not only are pion and kaon masses and decay constants
calculated quantities, predicted from the $q\bar q$ substructure, but once we
formulate the incorporation of Shore's generalization within the bound-state
DS approach, it will become obvious that also these four $\eta$ and $\eta'$
decay constants {\it and} their masses $M_\eta$ and $M_{\eta'}$ come out
as pure predictions.
Such a phenomenological analysis, complementary to Shores, motivates us
to formulate and perform the treatment based on Shore's generalization,
instead of the original WV relation (or fitting the anomalous $\eta_0$
mass shift) as in our earlier references \cite{Klabucar:1997zi,Kekez:2000aw,Kekez:2001ph,Kekez:2005ie,Horvatic:2007qs}.

Adding Eqs. (\ref{eq:bn}) and (\ref{eq:bp}), one gets the relation
\begin{eqnarray}
(f^0_{\eta'} )^2 M_{\eta'}^2  &+& (f^0_{\eta} )^2 M_\eta^2 
+ (f^8_{\eta})^2 M_\eta^2 \nonumber
\\
&+& (f^8_{\eta'})^2 M_{\eta'}^2 - 2f_K^2 M_K^2 = 6A
\end{eqnarray}
which is the analogue of the standard WV formula (\ref{WittenVenez}), 
to which it reduces in the large $N_c$ limit where $A\to \chi_{\rm YM}$, 
the $f^0_{\eta '}, f^8_{\eta }, f_K \to f_\pi$ limit, and the limit of
vanishing subdominant decay constants (since $\eta$ and $\eta'$ are 
dominantly $\eta_8$ and $\eta_0$, respectively), i.e.,
$f^0_{\eta }, f^8_{\eta'} \to 0$. However, we will need to use
not just this single equation, but the three equations 
(\ref{eq:bn})-(\ref{eq:bp}) from Shore's generalization.

These four $\eta$ and $\eta'$ decay constants are often 
parameterized in terms of two decay constants, $f_8$ and $f_0$,
and two mixing angles, $\theta_8$ and $\theta_0$: 
\begin{equation}
\label{def:f8etaf0eta}
f^8_\eta = \cos\theta_8\, f_8~, \qquad \,\,  f^0_\eta = -\sin\theta_0\, f_0~,
\end{equation}
\begin{equation}
f^8_{\eta'} = \sin\theta_8\, f_8~, \qquad   f^0_{\eta'} = \cos\theta_0\, f_0~.
\label{def:f8eta'f0eta'}
\end{equation}
This is the so-called two-angle mixing scheme, which shows explicitly 
that it is inconsistent to assume that the mixing of the decay constants
follows the pattern (\ref{MIXtheta}) of the mixing of the states 
$\eta_8$ and $\eta_0$
\cite{Gasser:1984gg,Schechter:1992iz,Leutwyler98,KaiserLeutwyler98,FeldmannKrollStech98PRD,FeldmannKrollStech99PLB,Feldmann99IJMPA}.

The advantage of our model is that, as we shall see, we are able to calculate
the $f_8$ and $f_0$ parts of the physical decay constants
(\ref{def:f8etaf0eta})-(\ref{def:f8eta'f0eta'}) from the $q\bar q$ substructure.
However, we cannot keep the full generality of Shore's approach, which 
allows for the mixing with the gluonic pseudoscalar operators, and 
therefore employs the definition \cite{Shore:2006mm,Shore:2007yn} 
of the decay constants which, in general, due to the gluonic contribution, 
differs from the following standard definition 
through the matrix elements of the axial currents $A^{a\,\mu}(x)$:
\begin{equation}
\!\! \langle 0|A^{a\,\mu}(x)|P(p)\rangle = if^a_P\, p^\mu e^{-ip\cdot x},
\,\,a=8,0;\,\,P=\eta,\eta^\prime~.
\label{def2angSch}
\end{equation}
Nevertheless, Shore's definition \cite{Shore:2006mm,Shore:2007yn}
coincides with the above standard one in the 
non-singlet channel, where there cannot be any admixture of the pseudoscalar 
gluonic component.  Similarly, since our BS solutions (from Refs. 
\cite{Klabucar:1997zi,Kekez:2000aw,Kekez:2005ie,Horvatic:2007qs}) 
are the pure $q\bar q$ states, without any gluonic components, using Shore's
definition would not help us calculate the gluon anomaly influence on the 
decay constants. We thus employ the standard definitions (\ref{def2angSch}), 
also used by, e.g., Gasser, Leutwyler, and Kaiser 
\cite{Gasser:1984gg,Leutwyler98,KaiserLeutwyler98},
as well as by Feldmann, Kroll, and Stech (FKS)
\cite{FeldmannKrollStech98PRD,FeldmannKrollStech99PLB,Feldmann99IJMPA}.

In Eqs. (\ref{def:f8etaf0eta})-(\ref{def:f8eta'f0eta'}), the angles are
chosen so \cite{Feldmann99IJMPA} that $\theta_8 = \theta_0 = \theta = 0$ 
in the limit of the exact SU(3) flavor symmetry, since only then there 
are just two decay constants, purely octet $f^8_{\eta} = f_8$ and purely
singlet $f^0_{\eta'} = f_0$, while the off-diagonal decay constants
vanish, $f^0_{\eta} = 0 = f^8_{\eta'}$, in this limit.
Otherwise, all four decay constants (\ref{def2angSch}) are different from 
zero due to the breaking of the SU(3) flavor symmetry, since this leads to 
$\theta \neq 0$ and gives both $\eta$ and $\eta'$ the both components 
$\eta_8$ and $\eta_0$. [In the parameterization 
(\ref{def:f8etaf0eta})-(\ref{def:f8eta'f0eta'}), the angles $\theta_8$
and $\theta_0$ differ from $\theta$ since also 
$\langle 0|A_\mu^8|\eta_0\rangle\neq 0 \neq \langle 0|A_\mu^0|\eta_8\rangle$.]
Thus, although not $\eta_8$ but $\eta_0$ couples to the gluon anomaly,  
the octet-chanel constants $f^8_{\eta}$ and $f^8_{\eta'}$ are influenced by the 
gluon anomaly through its interplay with the SU(3) flavor symmetry breaking
[similarly to the anomalous mass matrix (\ref{M2AqqX}) having nonvanishing
88, 08 and 80 elements when $X \neq 1$].

Equivalently to $f^0_{\eta'}, f^8_{\eta }, f^0_{\eta }$, and $f^8_{\eta'}$, 
defined by Eq. (\ref{def2angSch}), one has four related but different constants
$f^{\mathrm{NS}}_{\eta'}, f^{\mathrm{NS}}_{\eta }, f^\mathrm{S}_{\eta }$, and $f^\mathrm{S}_{\eta'}$, if instead 
of octet and singlet axial currents ($a=8,0$) in Eq. (\ref{def2angSch}) 
one uses the nonstrange-strange axial currents ($a=\mathrm{NS},\mathrm{S}$)
\begin{eqnarray}
A_\mathrm{NS}^{\mu}(x) &=&  \frac{1}{\sqrt{3}} A^{8\,\mu}(x)
                    + \sqrt{\frac{2}{3}} A^{0\,\mu}(x)
\nonumber
\\
                &=& \frac{1}{2} \left[
    \bar{u}(x)\gamma^\mu\gamma_5 u(x)
    +
    \bar{d}(x)\gamma^\mu\gamma_5 d(x)
                \right] ,
\end{eqnarray}
\begin{equation}
A_\mathrm{S}^{\mu}(x) = - \sqrt{\frac{2}{3}} A^{8\,\mu}(x)
                   + \frac{1}{\sqrt{3}} A^{0\,\mu}(x)
= \frac{1}{\sqrt{2}}\bar{s}(x)\gamma^\mu\gamma_5 s(x)~.
\end{equation}
The relation between the two equivalent sets is thus
\begin{equation}
\left[ \begin{array}{cc}
f^\mathrm{NS}_\eta &
        f^S_\eta
\\
f^\mathrm{NS}_{\eta^\prime} &
        f^S_{\eta^\prime}
       \end{array}
\right]
=
\left[ \begin{array}{cc}
f^8_\eta &
        f^0_\eta
\\
f^8_{\eta^\prime} &
        f^0_{\eta^\prime}
       \end{array}
\right]
\left[ \begin{array}{cc}
\frac{1}{\sqrt{3}} & -\sqrt{\frac{2}{3}}
\\
\sqrt{\frac{2}{3}} & \frac{1}{\sqrt{3}}
       \end{array}
\right]~.
\label{TwoMixingAngles:sns-etatap}
\end{equation}
Of course, this other quartet of $\eta$ and $\eta'$ decay constants
can also be  parameterized in terms of other two constants and two other
mixing angles:
\begin{equation}
\label{def:fNSetafSeta}
f^\mathrm{NS}_\eta = \cos\phi_\mathrm{NS}\, f_\mathrm{NS}~, \qquad \,\,  f^\mathrm{S}_\eta = -\sin\phi_\mathrm{S}\, f_\mathrm{S}~,
\end{equation}
\begin{equation}
f^\mathrm{NS}_{\eta'} = \sin\phi_\mathrm{NS}\, f_\mathrm{NS}~, \qquad  f^\mathrm{S}_{\eta'} = \cos\phi_\mathrm{S}\, f_\mathrm{S}~,
\label{def:fNSeta'fSeta'}
\end{equation}
where $f_\mathrm{NS}$ and $f_\mathrm{S}$ are given by the matrix elements
\begin{equation}
\langle 0| A_\mathrm{NS}^{\mu}(x) |\eta_\mathrm{NS}(p)\rangle
=
i f_\mathrm{NS}\, p^\mu e^{-ip\cdot x}~,
\end{equation}
\begin{equation}
\langle 0| A_\mathrm{S}^{\mu}(x) |\eta_\mathrm{S}(p)\rangle
=
i f_\mathrm{S}\, p^\mu e^{-ip\cdot x}~,
\end{equation}
while $\langle 0| A_\mathrm{NS}^{\mu}(x) |\eta_\mathrm{S}(p)\rangle = 0
= \langle 0| A_\mathrm{S}^{\mu}(x) |\eta_\mathrm{NS}(p)\rangle$.

In the $\mathrm{NS}$-$\mathrm{S}$ basis, it is possible to recover a scheme with a single mixing
angle $\phi$ through the application of the Okubo-Zweig-Iizuka (OZI) rule 
\cite{FeldmannKrollStech98PRD,FeldmannKrollStech99PLB,Feldmann99IJMPA}. 
For example, $f_\mathrm{NS} f_\mathrm{S} \sin(\phi_\mathrm{NS}-\phi_\mathrm{S})$ 
differs from zero just by an OZI-suppressed term \cite{Feldmann99IJMPA}. 
Neglecting this term thus implies $\phi_\mathrm{NS} = \phi_\mathrm{S}$.
(Refs. 
\cite{FeldmannKrollStech98PRD,FeldmannKrollStech99PLB,Feldmann99IJMPA}
denote $f_\mathrm{NS}, f_\mathrm{S}, \phi_\mathrm{NS}, \phi_\mathrm{S}$ by, respectively,
$f_q, f_s, \phi_q, \phi_s$.)
In general, neglecting the OZI-suppressed terms, i.e., 
application of the OZI rule, leads to the so-called FKS scheme
\cite{FeldmannKrollStech98PRD,FeldmannKrollStech99PLB,Feldmann99IJMPA},
which exploits a big practical difference between the (in principle
equivalent) parameterizations (\ref{def:f8etaf0eta})-(\ref{def:f8eta'f0eta'}) 
and (\ref{def:fNSetafSeta})-(\ref{def:fNSeta'fSeta'}):
while $\theta_8$ and $\theta_0$ differ a lot from each other and
from the octet-singlet {\it state} mixing angle 
$\theta \approx (\theta_8 + \theta_0)/2$, 
the $\mathrm{NS}$-$\mathrm{S}$ decay-constant mixing angles are very close to each other
and both can be approximated by the state mixing angle:
$\phi_\mathrm{NS} \approx \phi_\mathrm{S} \approx \phi$. Therefore one
can deal with only this one angle, $\phi$, and express
the physical $\eta$-$\eta'$ decay constants as
\begin{equation}
\! \! \left[ \begin{array}{cc}
f^8_\eta &
   f^0_\eta
\\
f^8_{\eta^\prime} &
   f^0_{\eta^\prime}
       \end{array}
\right]
\! = \!
\left[ \begin{array}{cr}
f_\mathrm{NS} \cos\phi & -f_\mathrm{S} \sin\phi 
\\
f_\mathrm{NS} \sin\phi &  f_\mathrm{S} \cos\phi
       \end{array}
\right]
\!
\left[ \begin{array}{cc}
\frac{1}{\sqrt{3}} & \sqrt{\frac{2}{3}}
\\
-\sqrt{\frac{2}{3}} & \frac{1}{\sqrt{3}}
       \end{array}
\right].
\label{matrixEqLast}
\end{equation}
This relation is valid also in our approach, where $\eta$ and $\eta'$ are 
the simple $\eta_\mathrm{NS}$-$\eta_\mathrm{S}$ mixtures (\ref{MIXphi}).
The FKS relations 
\cite{FeldmannKrollStech98PRD,FeldmannKrollStech99PLB,Feldmann99IJMPA}
\begin{equation}
f_8 = \sqrt{\frac{1}{3} f_\mathrm{NS}^2
     + \frac{2}{3} f_\mathrm{S}^2 }~,
\label{f_8}
\quad
\theta_8 = \phi - {\arctan}\left(\frac{\sqrt{2} f_\mathrm{S}}
                                    {f_\mathrm{NS}} \right)~,
\end{equation}
\begin{equation}
\! f_0 = \sqrt{\frac{2}{3} f_\mathrm{NS}^2
     + \frac{1}{3} f_\mathrm{S}^2 }~,
\,\,\,\,
\theta_0 = \phi - \mbox{\rm arctan}\left(\frac{\sqrt{2} f_\mathrm{NS}}
                                    {f_\mathrm{S}} \right)~,
\label{f_0}
\end{equation}
equivalent to Eq. (\ref{matrixEqLast}), were also shown \cite{Kekez:2000aw}
to hold in our DS approach.

In our present DS approach, mesons are pure $q\bar q$
BS solutions, without any gluonium admixtures, which are 
prominent possible sources of OZI violations. 
Therefore, our decay constants are calculated quantities,
$f_\mathrm{NS}=f_{u\bar u}=f_{d\bar d}=f_\pi$ and $f_\mathrm{S}=f_{s\bar s}$,
in agreement with the OZI rule. 
Our DS approach is thus naturally compatible with the FKS scheme,
and we can use the $\eta$ and $\eta'$ decay constants 
(\ref{matrixEqLast}) with our calculated $f_\mathrm{NS}=f_\pi$ and 
$f_\mathrm{S}=f_{s\bar s}$ in Shore's equations (\ref{eq:bn})-(\ref{eq:bp}).

\begin{table}[b]
\begin{center}
\begin{tabular}{|c||c|c||c|c||c|c||}
\hline
Inputs:     & \multicolumn{2}{|c||}{from  Ref. \cite{Kekez:2000aw}}   & \multicolumn{2}{|c||}{from  Ref. \cite{Kekez:2005ie}}  & \multicolumn{2}{|c||}{from  Ref. \cite{Horvatic:2007qs}  } \\
\hline
$\chi_{\mbox{\rm\scriptsize YM}}^{1/4}$         & $175.7$ & $191$    & $175.7$ & $191$    & $175.7$  & $191$    \\
\hline
$M_\eta$                                & 485.7     & 499.8      & 482.8     & 496.7      & 507.0      & 526.2      \\
$M_{\eta^\prime}$                       & 815.8     & 931.4      & 818.4     & 934.9      & 868.7      & 983.2      \\
$\phi$                                  & $46.11^\circ$ & $52.01^\circ$  & $46.07^\circ$ & $51.85^\circ$  & $40.86^\circ$  & $47.23^\circ$  \\
\hline
$\theta_0$                              & $1.84^\circ$  & $7.74^\circ$   & $1.39^\circ$  & $7.17^\circ$   & $-6.69^\circ$  & $-0.33^\circ$  \\
$\theta_8$                              & $-17.90^\circ$& $-12.00^\circ$ & $-17.6^\circ$ & $-11.85^\circ$ & $-20.47^\circ$ & $-14.11^\circ$ \\
$f_0$                                   & 108.8     & 108.8      & 107.9     & 107.9      & 101.8      & 101.8      \\
$f_8$                                   & 122.6     & 122.6      & 121.1     & 121.1      & 110.7      & 110.7      \\
$f_\eta^0$                              & -3.5      & -14.7      & -2.6      & -13.5      & 11.9       & 0.6        \\
$f_{\eta^\prime}^0$                     & 108.8     & 107.9      & 107.9     & 107.1      & 101.1      & 101.8      \\
$f_\eta^8$                              & 116.7     & 119.9      & 115.4     & 118.5      & 103.7      & 107.4      \\
$f_{\eta^\prime}^8$                     & -37.7     & -25.5      & -37.6     & -24.9      & -38.7      & -27.0      \\
\hline
\end{tabular}
\end{center}
\caption{The results of the three DS models
obtained through Shore's equations (\ref{eq:bn})-(\ref{eq:bp})
for the two values of $\chi_{\rm YM}$ approximating $A$:
$(175.7\rm MeV)^4$ and $(191\rm MeV)^4$.
Columns 1 and 2: The results when the non-anomalous inputs for
Eqs. (\ref{eq:bn})-(\ref{eq:bp}), namely $M_\pi, M_K, f_\pi=f_\mathrm{NS},
f_{s\bar s}=f_\mathrm{S}$ and $f_K$, are taken from Ref. \cite{Kekez:2000aw},
which uses Jain--Munczek {\em Ansatz} interaction \cite{jain93b}.
Columns 3 and 4: The results for the non-anomalous
inputs from Ref. \cite{Kekez:2005ie}
using OPE-inspired interaction nonperturbatively
dressed by gluon condensates \cite{Kekez:2003ri}.
Columns 5 and 6: The results for the
inputs from Ref. \cite{Horvatic:2007qs}
using the separable {\em Ansatz} interaction \cite{Blaschke:2000gd}.
All masses and decay constants, as well as
$\chi_{\mbox{\rm\scriptsize YM}}^{1/4}$, are in MeV,
and angles are in degrees.
}
\label{DGMOR:tab:eta-etap-mixing-all3m-forArticle}
\end{table}

\section{Results and conclusions}
\label{ResultsAndConclusions}

All quantities appearing on the right-hand side of Eqs.  (\ref{eq:bn})-(\ref{eq:bp}), 
namely $M_\pi$, $M_K$, $f_\pi$, and $f_K$, are calculated in our DS approach 
\cite{Kekez:2000aw,Kekez:2005ie,Horvatic:2007qs} (for the three dynamical
models \cite{jain93b,Kekez:2003ri,Blaschke:2000gd}), {\it except} the full 
QCD topological charge parameter $A$. Since it is at present
unfortunately not yet known, we follow Shore and approximate it
by the Yang-Mills topological susceptibility $\chi_{\rm YM}$.

On the left-hand side of Eqs. (\ref{eq:bn})-(\ref{eq:bp}), 
the model results for $f_\mathrm{NS}=f_\pi$ and $f_\mathrm{S}=f_{s\bar s}$ 
and Eq.~(\ref{matrixEqLast}) reduce the unknown part of the four 
$\eta$ and $\eta'$ decay constants $f_\eta^0$, $f_{\eta^\prime}^0$, 
$f_\eta^8$, and $f_{\eta^\prime}^8$, down to the mixing angle $\phi$.
The three Shore's equations (\ref{eq:bn})-(\ref{eq:bp}) can then
be solved for $\phi$, $M_\eta$ and $M_{\eta^\prime}$, providing us
with the upper three lines of            
Table \ref{DGMOR:tab:eta-etap-mixing-all3m-forArticle}.
For each of the three different dynamical models
which we used in our previous DS studies
\cite{Klabucar:1997zi,Kekez:2000aw,Kekez:2005ie,Blaschke:2007ce,Horvatic:2007wu,Horvatic:2007qs},
these results are displayed
for $\chi_{\rm YM} = (175.7\, \rm MeV)^4$
as in Refs. \cite{Kekez:2005ie,Horvatic:2007qs}
and for $\chi_{\rm YM} = (191\, \rm MeV)^4$ \cite{DelDebbio:2004ns}
(adopted by Shore \cite{Shore:2006mm,Shore:2007yn}).
The lower part of the table, displaying various additional results,
is then readily obtained through Eq. (\ref{matrixEqLast}) and/or 
the useful relations (\ref{f_8})-(\ref{f_0}) which give $f_8$, $f_0$,
$\theta_8-\phi$ and $\theta_0-\phi$ in terms of $f_\mathrm{NS} =f_\pi$ 
and $f_\mathrm{S} = f_{s\bar s}$.
Thus, unlike the mixing angles, $f_0$ and $f_8$ do not result from solving 
of Eqs. (\ref{eq:bn})-(\ref{eq:bp}), but are the calculated predictions of a
concrete dynamical DS model, independently of Shore's equations.

For all three quite different (RGI \cite{jain93b,Kekez:2003ri} and
non-RGI \cite{Blaschke:2000gd}) dynamical models
which we used in our previous DS studies
\cite{Klabucar:1997zi,Kekez:2000aw,Kekez:2005ie,Blaschke:2007ce,Horvatic:2007wu,Horvatic:2007qs},
the situation with the results turns out to be rather similar.
Similar results from various models mean that the usage of Shore's
generalization in conjunction with the DS approach does not help one
to discriminate between various dynamical models and so draw conclusions
on the dynamics. This is not surprising, as it has been established 
\cite{ForExample,Alkofer:2000wg,Roberts:2000hi,Roberts:2000aa,Holl:2006ni,Fischer:2006ub}
that while a successful reproduction of static properties and other low-energy
meson phenomenology requires interaction modeling at low momenta, it is possible
to achieve a satisfactory description of low-energy phenomenology for many forms
of model interactions as long as their integrated strength at low momenta
($p^2 < 1$ GeV$^2$) is sufficient to achieve a realistic DChSB. On the other hand, 
this (similarity of our results from the very different models) has the
advantage that our conclusions further below are not sensitive to the changes
of the model dynamics.

The most conspicuous feature of our results is that $\eta$ and 
$\eta'$ masses are both much too low when the weighted
average $\chi_{\rm YM} = (175.7 \pm 1.5 \, \rm MeV)^4$ of Refs.
\cite{Lucini:2004yh,DelDebbio:2004ns,Alles:2004vi} is used, in contrast to
the results from the standard WV relation, displayed in Table \ref{3WV+Shore}.
If we single out just the highest of these values ($191 \, \rm MeV)^4$
\cite{DelDebbio:2004ns}), the masses improve somewhat. However,
other results are spoiled -- e.g., the mixing angle $\phi$
becomes too high to enable agreement with the experimental results on
$\eta, \eta' \to \gamma\gamma$ decays, which require $\phi \sim 40^\circ$
\cite{Kekez:2005ie}.

When we turn to the lower parts of Tables \ref{3WV+Shore} and
\ref{DGMOR:tab:eta-etap-mixing-all3m-forArticle}, where the
results for the $\eta$ and $\eta'$ decay constants, and the
corresponding two mixing angles $\theta_0$ and $\theta_8$,
are given, we notice a feature common to all our results,
as well as Shore's (also given in Table \ref{3WV+Shore}).
The diagonal ones, $f_{\eta^\prime}^0$ and $f_\eta^8$, are
all of the order of $f_\pi$, being larger by some 10\% to
30\%. The off-diagonal ones, $f_{\eta^\prime}^8$ and $f_\eta^0$,
are, on the other hand, in general strongly suppressed.
This is expected, as $\eta^\prime$ is mostly singlet,
and $\eta$ is mostly octet. 

To understand the dependence 
of the decay constants on the topological susceptibility 
$\chi_{\rm YM}$ (approximating $A$), it is important to note 
that our $f_0$, which in a full QCD bound-state calculation
would be influenced by the gluon anomaly, presently is not, 
since it is calculated (same as $f_8$) from the modeled
meson $q\bar q$ substructure relying on RLA. 
In Tables I and II one therefore sees no $\chi_{\rm YM}$-dependence of 
not only $f_8$, but also of $f_0$, since the difference between $f_0$ 
and $f_8$ is presently generated only by their different $\mathrm{NS}$
and $\mathrm{S}$ quark content. This feature is not only consistent with 
the FKS scheme, but is in fact a general characteristic of this scheme.
Namely, due to the neglect of OZI-violating contributions
\cite{Feldmann99IJMPA}, in the SU(3) flavor symmetry limit 
one would have $f_\mathrm{NS} = f_\mathrm{S}$ and $f_0 = f_8$.
(The DS approach will be able to obtain the 
gluon anomaly dependence of $f_0$ only when it manages to go beyond RLA,
which is presently achieved only with schematic, very simplified
$\delta$-function-type interactions \cite{Bhagwat:2007ha}.)
Thus, not only in the present DS calculation, but in fact in any 
application of the FKS scheme, the $\chi_{\rm YM}$-dependence of the 
four physical $\eta$-$\eta'$ decay constants 
(\ref{def:f8etaf0eta})-(\ref{def:f8eta'f0eta'})
stems exclusively from the
$\chi_{\rm YM}$-dependence of the mixing angles $\phi, \theta_8$ and 
$\theta_0$. Its origin, as explained in the previous section, is in 
the interplay of the anomaly with the flavor symmetry breaking. 
In fact, the FKS scheme is based on the assumption that the 
flavor symmetry breaking is significantly more important than 
the OZI-violating contributions (arising beyond RLA in DS approach).

The feature that may be surprising is that Shore's results
(which, to be sure, were obtained \cite{Shore:2006mm,Shore:2007yn}
in quite a different way from ours) are more similar to our results
obtained through the standard WV relation, than to our results
obtained through Shore's Eqs. (\ref{eq:bn})-(\ref{eq:bp}).

To summarize: the present paper has explored a modification of 
the DS treatments of the $\eta$-$\eta'$ complex employed in 
Refs. \cite{Klabucar:1997zi,Kekez:2000aw,Kekez:2001ph,Kekez:2005ie}.
In Refs. \cite{Klabucar:1997zi,Kekez:2000aw,Kekez:2001ph}, the
value of the anomalous $\eta_0$ mass shift was obtained by fitting, 
but Ref. \cite{Kekez:2005ie} improved the treatment by obtaining 
it from the lattice through the WV relation. A generalization of this
relation was recently proposed \cite{Shore:2006mm,Shore:2007yn}, 
and the purpose of the present paper is to tests the usage thereof 
in the bound-state, DS context, and compare the results with those
from the standard WV relation.

All in all, inspection and comparison of the results 
in Table \ref{DGMOR:tab:eta-etap-mixing-all3m-forArticle}
with the results (in Table \ref{3WV+Shore}) from the
analogous calculations but using the standard WV relation to construct
the complete $\eta$-$\eta'$ mass matrix, leads to the conclusion that
the DS approach with the standard WV relation (\ref{WittenVenez}) is 
phenomenologically more successful, yielding the masses closer to 
the experimental ones.  This may seem surprising,
but one must be aware that we do not
yet have at our disposal the full QCD topological charge parameter
$A$, and that we (along with Shore) had to use its lowest ${1}/{N_c}$
approximation, $\chi_{\rm YM}$.
This in general precludes a {\it consistently} improved $1/N_c$ treatment in
spite of the usage of Shore's relations. The problems with inconsistencies
in the $1/N_c$ counting may well cause spoiling of results, especially in an
approach such as ours, where the $\eta$ and $\eta'$ masses are not inputs,
but predicted quantities.
Our results thus add a new argument to the motivation for undertaking
lattice calculations proposed by Shore \cite{Shore:2007yn} and aimed
at proper finding the quantity $A$.
Also, we should recall from Sections \ref{INTRO} and
\ref{massMatrixAndWVrelation} that the very usage of the RLA
assumed that the anomaly is implemented on the level of
the anomalous mass only, as a lowest order $1/N_c$ correction
\cite{Klabucar:1997zi,Kekez:2000aw,Kekez:2005ie,Horvatic:2007wu,Horvatic:2007qs}.
Thus, with respect to the orders in ${1}/{N_c}$,
using Shore's generalization in the present formulation of our DS approach
may be less consistent than using the standard WV relation, which
may well be the cause of its lesser phenomenological success.

In spite of the lesser phenomenological success (than the standard WV relation) 
in the present context of bound-state DS calculations at zero temperature, the 
presently exposed usage of Shore's generalization will likely find its application 
at finite-temperature calculations in the DS context. Namely, there it may help 
alleviate the difficulties met due to the usage of the standard WV relation in 
the DS approach at $T>0$, as discussed in Ref. \cite{Horvatic:2007qs}.

{\bf Acknowledgments} 

D.H.~and~D.Kl. acknowledge the support of the project No.~119-0982930-1016 
of MSES of Croatia. D.Kl. also acknowledges the hospitality and support through
senior associateship of International Centre for Theoretical 
Physics at Trieste, Italy, where the present paper was started.  
D.Kl. also thanks the LIT of JINR for 
its hospitality in Dubna, Russia, in August 2007. D.Ke. acknowledges 
the support of the Croatian MSES project No.~098-0982887-2872.
Yu.K. thanks for support from Deutsche Forschungsgemeinschaft (DFG) 
under grant No. BL 324/3-1, and the work of D.B. was supported by 
the Polish Ministry of Science and Higher Education under contract 
No. N N202 0953 33.

\end{document}